\documentclass[12pt,preprint]{cmlapss}
\newcommand{\beq}{\begin{equation}}
\newcommand{\eeq}{\end{equation}}
\newcommand{\beqn}{\begin{eqnarray}}
\newcommand{\eeqn}{\end{eqnarray}}
\newcommand{\no}{\noindent}
\newcommand{\non}{\nonumber}
\begin{document}
       
\title{\bf On the Scaling and Spacing 
of Extra-Solar Multi-Planet Systems
}
\author{Li-Chin Yeh$^{1}$, 
Ing-Guey Jiang$^{2,3,4}$, 
Sridhar Gajendran$^{2}$ 
}
\affil{
{$^{1}$Institute of Computational and Modeling Science,}\\ 
{National Tsing-Hua University, Hsin-Chu, Taiwan}\\ 
{$^{2}$Institute of Astronomy,}
{National Tsing-Hua University, Hsin-Chu, Taiwan}\\
{$^{3}$Department of Physics,}
{National Tsing-Hua University, Hsin-Chu, Taiwan}\\
{$^{4}$Center for Informatics and Computation in Astronomy,}\\
{National Tsing-Hua University, Hsin-Chu, Taiwan}\\
}
\email{jiang@phys.nthu.edu.tw}

\begin{abstract}
We investigate whether certain extra-solar multi-planet systems
simultaneously follow the scaling and spacing rules 
of the angular-momentum-deficit model.
The masses and semi-major axes of exoplanets 
in ten multi-planet systems are considered.
It is found that GJ 667C, HD 215152, HD 40307, and Kepler-79
systems are currently close to configurations
of the angular-momentum-deficit model.
In a gas-poor scenario, GJ 3293, HD 141399, and HD 34445 systems 
are those which 
had a configuration of the angular-momentum-deficit model 
in the past and get scattered away due to post gaseous effects. 
In addition, no matter in gas-free or gas-poor scenario, 
55 Cnc, GJ 876, and WASP-47 systems do not follow
the angular-momentum-deficit model.   
Therefore, our results reveal important formation histories
of these multi-planet systems.

\end{abstract}

\keywords{planetary systems $-$ stars: individual}

\section{Introduction}
It is well known that the relative spacing of most planets in
the Solar System approximately
follow an empirical rule, i.e. the Titus-Bode Rule (TBR). 
Although the physical mechanism of this rule was unclear, it 
successfully predicted the existence of Uranus and Ceres.
The discoveries of these celestial bodies were remarkable 
and regarded as a great success of TBR.
The significance of the TBR for the distribution of 
the planets in the Solar system was later statistically investigated in 
Lynch (2003).

The above success motivated some theoretical 
investigations on possible
mechanisms which could explain the planet spacing.  
This work was based on the physics of planetary formation.
For example, Rawal (1986) studied the spacing relations through
the formation and evolution of gaseous rings.  
Patton (1988) tried to obtain the planet spacing
through an algorithm derived from the
principle of least action interaction. 
Though Patton (1988) did not provide
an analytic formula,
the numerical results of planet spacing 
were very close to the TBR.
This implied that the least action interaction
could be a physical reason that
the TBR gave a good approximation for 
planet and satellite spacing in the Solar System.  
In addition,
Hayes \& Tremaine (1998) employed a stability condition which was
based on Hill radii to address the spacing of the Solar system, 
and concluded that the planets are not randomly 
but regularly spaced in a stable system.
Later, Griv \& Gedalin (2005) found that a theory of
disk instability could lead to another rule of  
planet spacing, which is different though similar to 
the TBR.

More recently, Christodoulou \& Kazanas (2017) considered the 
geometric TBR for the semi-major axes of planetary orbits. 
They interpreted it in terms of the work done in the 
gravitational field of the central star 
by particles whose orbits are perturbed around each planetary orbit.
Further, Laskar (2000) and Laskar \& Petit (2017)
developed a model of planet formation based on 
an angular-momentum-deficit theory.
Their model starts from a phase 
that the disk is composed of planet embryos and planetesimals
after the gas is depleted.
Two main analytic laws are obtained and presented in their papers.
The first one gives a scaling relation between the 
period-ratio and mass-ratio of adjacent planets 
(the scaling rule hereafter).
The second one leads to the spacing between adjacent planets
(the spacing rule hereafter). This spacing rule is equivalent to
the TBR for a particular mass distribution of planetesimal
disk, and thus can be regarded as a generalized TBR.  
  
The discoveries of extra-solar planets (exoplanets) 
gave excellent opportunities to further examine the
TBR.  For example, the 55 Cancri planetary system
was studied by Poveda \& Lara (2008) and Cuntz (2012). The
TBR was employed to compare its predictions 
with the planet spacing.
Interestingly,  Cuntz (2012) further predicted possible new
planet candidates in the 55 Cancri planetary system through
the TBR, and claimed one of them to be located in 
the outskirts of the habitable zone.
Bovaird \& Lineweaver (2013) studied those 
extra-solar planetary systems
with higher multiplicity. They found that these planetary systems 
do not exactly follow the TBR. Nevertheless, they    
obtained a new two-parameter relation empirically. 
As a test of this new relation, 
Huang \& Bakos (2014) then used Kepler data 
to search for possible new planets predicted by that relation. 
They successfully identified five planetary candidates around
predicted positions.
However, among most systems in their study,
there were no signals of new planets.
Moreover, Aschwanden (2018) developed a rule of quantized harmonic ratios
and applied to the planet spacing of extra-solar planetary system
and also moon spacings in the Solar system.  
Finally, Pletser (2019) found that the spacing of the above systems 
could be related to Fibonacci numbers. 

On the other hand, because the mass and orbital period 
are the most important parameters, 
there have been many studies about the distributions of exoplanets
in the period-mass plane  
(Zucker \& Mazeh 2002, Tabachnik \& Tremaine 2002, 
Jiang et al. 2006).
Additionally, the coupled period-mass functions were first 
explored in Jiang et al. (2007, 2009), and then further investigated
with proper treatments of the selection effect in Jiang et al. (2010).
Moreover, Jiang et al. (2015) studied the 
period-ratio-mass-ratio correlation
of adjacent planet pairs in 166 multiple planetary systems.
A moderate correlation between the period-ratio and mass-ratio
was found with a correlation coefficient 0.5303.

This correlation between adjacent planets
strongly indicates that planets might follow 
the scaling or spacing rule.
It motivates us to investigate whether there 
are any known multi-planet systems that satisfy 
both the scaling and spacing rules in the angular-momentum-deficit (AMD) 
model, which has a clear physical foundation and 
provides analytic expressions. 
                        
The main expressions of the AMD model are summarized in Section 2.
The employed data of planetary systems is described in Section 3.
The data-model fitting results are in Section 4. 
We propose a scenario with gaseous effects to explain the data-model deviations 
and do another fitting in Section 5. We convey conclusions in Section 6. 

\section{The AMD Model}

Based on Eq. (10) in Laskar \& Petit (2017), 
the total angular momentum of a planetary system would 
have its maximum value when all bodies are moving in circular orbits 
in a coplanar system. For a realistic system, the bodies' orbits have  
different eccentricities and inclinations, and thus the total angular
momentum is smaller.
The AMD is defined to be the difference between
the maximum and the real value of total angular momentum. The AMD could 
evolve with time when there are collisions between bodies, i.e. 
planetesimals or embryos. In fact, the AMD was used to study 
the evolution of inner Solar system (Chambers 2001).  

During the stage of planetary formation, collisions between planetesimals 
and embryos could happen continuously in the disk until the system is 
settled and becomes more stable.
A planetary system is called AMD-stable
if its AMD is not sufficient to allow for further collisions.  
Laskar \& Petit (2017) derived the predicted planetary distribution
in an AMD-stable system. Their results lead to a scaling rule for
adjacent planets, and a spacing rule for the planets in the system.

Using semi-major axis $a$ as the variable, the mass distribution 
of a planetesimal disk is set to be
\beq
\frac{dm}{da}\equiv \rho(a)= \rho_0 a^p,
\eeq
where $m$ is the planetesimal mass as a function of $a$,
$\rho(a)$ is the linear mass density as a function of $a$,
$\rho_{0}$ is a constant, and $p$ is the power index.  
Depending on the disk structure, 
the value of $p$ could be negative, zero, or positive.
According to Laskar \& Petit (2017), for a given $p$, the disk surface density 
is proportional to $r^{p-1}$. 
For example, when $p=0$, it corresponds to a surface density 
proportional to $r^{-1}$.
Laskar (2000) discussed the cases when $p$=0, -1/2, -1, and -3/2.
In this paper, $p$ is allowed to be any real number, which 
would be determined through the data-model fitting.
However, for those models with $p>1$, their disk surface density
would increase with the radius $r$, and thus are considered as
unphysical models.

The scaling rule for a pair of consecutive planets
is stated as: 
\beq
\frac{m_i}{m_o} 
=\left( {\frac{a_i}{a_o}} \right) ^{\frac{1}{2}+ \frac{2}{3}p},
\eeq 
or equivalently
\beq
\ln\left(\frac{m_o}{m_i}\right)=
\left[\frac{1}{2}+\frac{2}{3}p\right] 
\ln\left(\frac{a_o}{a_i}\right),
\label{eq:ln_ma}
\eeq
where $m_o$ and  $m_i$ are the masses of the outer and inner planets and
$a_o$ and $a_i$ are the semi-major axes of the outer and inner planets,
respectively.

On the other hand, the spacing rule is separated into two cases 
in terms of the power index $p$.  
When $p \neq -3/2$, the semi-major axis $a$ of the $n$-th planet
satisfies 
\beq
a^b = a_0^b 
+ b D_0 n,
\label{eq:na}
\eeq
with
\beq
b\equiv \frac{2 p + 3}{6},
\label{eq:b}
\eeq
\beq
D_0\equiv \left(\frac{4C}{\rho_0\sqrt{GM_s}}\right)^{1/3},
\eeq
where $a_0$ is a constant, $C$ is the AMD,
$G$ is the gravitational constant,
and $M_s$ is the mass of the central star of the multi-planet system.  
When $p = -3/2$, the relation is
\beq
\ln (a) =  \ln (a_0) + D_0 n.
\label{eq:lna}
\eeq

In this paper, we will examine whether those planets in 
multi-planet systems 
satisfy both scaling and spacing rules of the AMD model.
These results would lead to important information on 
planet formation in multi-planet systems.

\section{The Exoplanet Data}
To obtain suitable samples for our study, 
the exoplanet catalog in http://exoplanet.eu/catalog/ 
was utilized.
The conditions of selecting multi-planet systems were set as:\\
(a) the number of planets is three or larger;\\
(b) the values of minimum masses and 
the corresponding error-bars are available;\\
(c) the values of semi-major axes and the corresponding error-bars 
are available;\\
(d) the error of minimum mass cannot be larger than 
   the minimum mass itself;\\
(e) the error of semi-major axis cannot be larger than 
   the semi-major axis itself.\\

In the end, ten planetary systems are chosen as the samples.
The names, the numbers of planets, and the references 
of these systems are shown in Table 1.\\

{\centerline {\bf Table 1}}
\begin{center}
\begin{tabular}{|c|c|c|c|}\hline
System ID& Name & Planet Number & Reference\\ \hline
1        &  55 Cnc  & 5 & Nelson et al. (2014)\\ \hline
2        &  GJ 3293 & 4 & Astudillo-Defru et al. (2017)\\ \hline
3        &  GJ 667C & 6 & Anglada-Escude et al. (2013)\\ \hline
4        &  GJ 876  & 4 & Jenkins et al. (2014)\\ \hline
5        &  HD 141399  & 4 & Hebrard et al. (2016)\\ \hline
6        &  HD 215152  & 4 & Delisle et al. (2018)\\ \hline
7        &  HD 34445  & 6 & Vogt et al. (2017)\\ \hline
8        &  HD 40307  & 6 &  Diaz et al. (2016)\\ \hline
9        &  Kepler-79 & 4 & Jontof-Hutter et al. (2014) \\ \hline
10       &  WASP-47   & 4 & Weiss et al. (2017) \\ \hline
\end{tabular}
\end{center}

The minimum masses, semi-major axes, and their corresponding errors 
regarding the 
planets in the above ten multi-planet systems, originally from
the references in Table 1, are listed in Appendix A. 
Because the orbital inclinations of many planets in the above ten systems 
are not known, the confirmed planetary masses are not available in general.
Nevertheless, as presented in the previous section, 
only mass ratios of two adjacent planets are involved 
in the AMD model, we thus use the values of minimum mass to 
represent the real masses in this paper. This leads to correct 
mass ratios under the assumption that these multi-planet 
systems are coplanar systems. This shall be a good approximation here.
In addition, the values of planetary mass hereafter means
the values of minimum masses shown in Appendix A.  

\section{A Gas-Free Scenario}
The scaling rule and spacing rule of the AMD model 
were obtained based on 
the assumption that the gas was depleted during the early 
stage of planet formation. 
In a gas-free environment, the planetesimals and embryos 
dynamically involved to a settled configuration.
If that process takes place and such a  
configuration exists, the planets shall follow 
what the AMD model predicts.
Thus, what we aim to investigate here is whether the 
masses and semi-major axes of exoplanets are governed
by the scaling and spacing rules. 

Ideally, the data can be taken to fit both rules simultaneously.
However, the spacing rule is divided into two different equations 
based on the value of the power index $p$.
Thus, we first examine whether 
the planetary masses and semi-major axes 
follow the scaling rule, and through the fitting, 
we obtain the best-fit $p$. Using this value of $p$, 
the spacing rule is then fitted with data accordingly.

In order to simplify the notation of the scaling rule in Eq.(3),  
we set 
$y=\ln\left({m_o}/{m_i}\right)$ and 
$x=\ln\left({a_o}/{a_i}\right)$.
Thus, the scaling rule, $y= (1/2 + 2p/3) x$, is a straight line 
passing through the origin.
For the corresponding observational data,
we use $y_j$ and $x_j$, $j=1,2,...,N$, 
where $N$ is the number of pairs. 
For example, $y_1$ is the logarithm of 
planetary mass ratio of innermost adjacent pair,
and $x_1$ is the logarithm of the ratio of semi-major axes  
of the innermost adjacent pair.

The values of $y_j$ and $x_j$ can be calculated 
from observational values 
of planetary masses and semi-major axes listed in Appendix A.
Through the method of error propagation (see Appendix B),
the error bars of both $y_j$ and $x_j$ can be determined.

For the data-model fitting, the $\chi^2$ minimization  
is employed (Press et al. 1992). 
Through a Markov Chain Monte Carlo (MCMC) sampling
to minimize the $\chi^2$ function, 
the reduced $\chi^2$ which measures the agreement
between the data and the rule is determined. 
In this paper, the {\it emcee} code (Foreman-Mackey et al. 2013)  
is chosen as the tool for the MCMC sampling.

For convenience, we define $\beta \equiv 1/2 + 2p/3$, and 
thus the scaling rule is simply $y= \beta x$.
During the MCMC sampling of $\chi^2$ minimization,
$y$ is the fitting function, $x$ is the variable, and $\beta$
is the fitting parameter. Through the $\chi^2$ minimization, 
the $\beta$ and the corresponding reduced $\chi^2$, i.e. $\chi^2_{r1}$,
would be determined. 
Then, the best-fit parameter $\beta$ leads to the
best-fit value of $p$. 
Once the best-fit power-index $p$ is obtained, 
the value $p$ is substituted into the spacing rule. 
Then, the spacing rule is also fitted in a similar way.
The best-fit parameters $D_0$, $a_0^b$, 
and the corresponding reduced $\chi^2$, i.e. $\chi^2_{r2}$,
can be obtained. 
 
All the above results are presented in Table 2. 
For System 1 and 10, it is found that
$p$ is larger than one and $a_0^b$ is negative,
and thus the corresponding models are unphysical.
Fig. 1-2 show the observational data with error bars,
and the best-fit straight line of the scaling rule  
and the spacing rule, respectively.
In addition, the values of reduced $\chi^2$ indicate
whether the fittings are satisfactory. For a good fitting,
the reduced $\chi^2$ shall be less than one or two
(Wall \& Jenkins 2003). 
From Table 2, 
it is clear that the $\chi^2_{r1}$ of System 3, 6, 9
are all smaller than one, the $\chi^2_{r1}$ of System 8
is around two, and the $\chi^2_{r1}$ of the rest
indicate bad fittings.
As shown in Fig. 1, System 3, 6, 8, 9 are consistent with 
the scaling rule approximately.

For the spacing rule, we only need to discuss the results
of System 3, 6, 8, 9 because the other systems already fail to fit
the scaling rule.
It is found that the $\chi^2_{r2}$ of System 3, 6, 9
are smaller than one, and the $\chi^2_{r2}$ of System 8
is around two. They follow the spacing rule as presented in Fig. 2. 
Considering both scaling and spacing rules, 
we conclude that System 3, 6, 8, 9, i.e. GJ 667C, HD 215152, HD 40307, 
Kepler-79 are consistent with the AMD model.
It is noted that their corresponding AMD models are all
phyisically acceptable.\\
  
{\centerline {\bf Table 2}}
\begin{center}
\begin{tabular}{|c|c|c|c|c|c|c|}\hline
ID & $p$ & $\chi^2_{r1}$ & $D_0$ & $a_0^b$ & $a_0$  & $\chi^2_{r2}$ \\
   &   &             &(AU$^b$)&(AU$^b$)& (AU) &          \\\hline 
1 &$1.16_{-0.10}^{+0.10}$ & 63.88 &$0.27_{-0.003}^{+0.003}$ &
 $-0.23_{-0.0053}^{+0.0053} $ &--&   193.33 \\ \hline
2 & $1.21_{-0.33}^{+0.33}$ & 56.46 & ${0.098}_{-0.018 }^{+0.018}$ & 
${0.0052}_{-0.039}^{+0.039}$ &  0.0029 & 1.03 
\\ \hline
3 &$-1.23_{-0.91}^{+0.89}$ & 0.20 & $0.37_{-1.01}^{+1.02}$ & $0.75_{-0.47}^{+0.46}$ & 
0.038  & $0.0013 $ \\ \hline  
4&$-1.26_{-0.06}^{+0.06}$ & 6714.12 & $0.58_{-0.18}^{+0.18}$ 
& $0.74_{- 0.048}^{+0.048} $ & 0.021 & 0.66 \\ \hline
5 &$-0.39_{-0.19}^{+0.20}$ & 38.65 & $0.70_{-0.11}^{+0.11}$ 
& $ 0.40_{- 0.080}^{+0.080}$  & 0.082 &4.06 \\ \hline
6 & $0.15 _{-1.74}^{+1.75}$ & 0.27 & ${0.085}_{-0.30}^{+0.30}$ 
& $0.15_{- 0.43}^{+0.43}$  & 0.030& 0.0056  \\ \hline 
7 &$-0.46_{-0.36}^{+0.38}$ & 15.01 & $0.52_{-0.086}^{+0.087}$ 
& $ 0.39_{-0.094 }^{+0.093}$ & 0.067 & 0.70 \\ \hline
8 &$0.34_{-0.28}^{+0.28}$ &1.80 & $0.16_{-0.021}^{+0.021}$ 
& ${0.020}_{-0.046}^{+0.046}$ & 0.0017 &  1.87\\ \hline
 9 &$-1.78_{-1.58}^{+1.57}$ & 0.20 &  $0.45 _{-4.34}^{+4.33}$
 & $1.26_{-1.35}^{+1.35} $ & 0.085 &$9.96 \times 10^{-5}$  \\ \hline
10 & $4.10_{-0.92}^{+1.46}$  & 13.42 &${0.0023}_{-0.0026}^{+0.0026}$& 
${-0.0038}_{- 0.0052}^{+0.0051} $& -- & 1.74 \\ \hline
\end{tabular}
\end{center}

\section{A Gas-Poor Scenario}
Here we consider a situation that the gaseous disk is not 
completely depleted 
while the planetesimals and planetary embryos
start to grow through collisions. 
Because the collisions and interactions between objects 
are completely random, the net result could be very close
to what the AMD model predicts.
Statistically,
the presence of a small amount of gas might 
only make the process described in the AMD model slower.
Therefore, there could be a phase that the planetary system
is consistent with a configuration described by the AMD model.
After planets settle to that configuration
with particular masses and semi-major axes,
the existence of gas would cause
two further slow processes described below.

Firstly, 
because planets have grown to certain sizes,
their gravitational force can attract the nearby gas efficiently. 
The gas would accrete onto the planets and increase the 
planetary masses. Once this process finishes, 
the planets are surrounded by gaps. They would stop growing 
and reach the final masses as observed today.
Secondly,
after the planetary masses are settled and the gaps are opened, 
planets might migrate slightly
due to the torque caused by the gaseous disk.
Their current semi-major axes are thus different from 
what the AMD model predicts.
Thus, the system would change from 
the AMD-model configuration 
due to post gaseous effects.
This could produce the observed deviation 
from the AMD model. 

Therefore, we investigate 
whether there was such a AMD-model phase.
In order to move backwards in time for the above mentioned two processes, 
we first allow the semi-major axes of the planets to 
be different from the current observed values, and inspect
which combination could improve the fitting 
of the scaling and spacing rules. 
We then assign several masses, 
which are the same or less than the observed values, 
to each planet of a multi-planet system and examine
which combination leads to the best-fit configuration.

There is no general rule how far each planet could be deviated from
the current orbit in the past.
For our purpose, it could be good enough
if we approximately allow one-third of the current 
spacing between two neighbors to be the  
maximum semi-major-axis difference.
For a given multi-planet system,
the innermost planet, i.e. the 1st planet, has 
a current semi-major axis $a_1$,  the 2nd planet has
a current semi-major axis $a_2$, and the $i$th
planet has a current semi-major axis $a_i$,
$i=3,\cdots,n$, where $n$ is the total number of planets
in this system.
The five values of semi-major axis assigned to 
the innermost planet are 
$a_1 - a_1/3$,  $a_1 - a_1/3 + a_1/6$,
$a_1$, $a_1 + a_{2,1}/6$, $a_1 + a_{2,1}/3$,
where  $a_{2,1}\equiv a_2 - a_1$.
The five values of semi-major axis assigned to 
the 2nd planet are 
$a_2 - a_{2,1}/3$, $a_2 - a_{2,1}/3 + a_{2,1}/6$, 
$a_2$, $a_2 + a_{3,2}/6$, $a_2 + a_{3,2}/3$,
where  $a_{3,2}\equiv a_3 - a_2$.
The five values of semi-major axis  
assigned to the the $i$th planet, $i=3,\cdots,n-1$
can be worked out similarly.
Finally, the five values of semi-major axis assigned to 
the $n$th planet, i.e. the outermost planet, are
$a_n - a_{n,n-1}/3$,  
$a_n - a_{n,n-1}/3 + a_{n,n-1}/6$, 
$a_n$, $a_n + a_{n,n-1}/6$, $a_n + a_{n,n-1}/3$,
where $a_{n,n-1}\equiv a_n - a_{n-1}$.

Each planet is assigned one of the above semi-major axes 
and the combination is considered as one of the possible past 
configuration of planets. 
The error bars of semi-major axes, planetary masses, 
and the error bars of masses
are set to be the observed values here.
For each of the possible configuration, the reduced $\chi^2$
and the best-fit power-index $p$
of the scaling rule are determined.
After all reduced $\chi^2$ of different combinations are obtained,
the configuration with the smallest reduced $\chi^2$
is the most likely one at this stage.

Secondly, in order to consider the mass accretion process back in time,
we now allow each planet to have five possible values of mass. 
The maximum is set to be $m+m_{+}$, where $m$ is the current observed mass
and $m_{+}$ is the upper bound of the error bar (Appendix A).
The minimum is set to be $(m+m_{+})/3$.
Then, three more values are set between them with uniform gaps.

Similarly, each planet is assigned one of the above masses
and the combination is considered as one of the possible past 
configuration of planets. Their semi-major axes are those 
from the most likely
choices we just determined.
Through exactly the same process, 
the configuration with the smallest reduced $\chi^2$, which is 
set as $\chi^2_{r1}$, 
is regarded as the past configuration of the planetary system
in this gas-poor scenario.
With the above best-fit power-index $p$, 
the spacing rule is fitted in a similar way.
The best-fit parameters $D_0$, $a_0^b$, 
and the corresponding $\chi^2_{r2}$ are therefore determined.

For all considered planetary systems, the 
corresponding parameters of past configurations
are presented in Table 3.
For System 1, it is found that $a_0^b$ is negative;
for System 10, $p$ is larger than one and $a_0^b$ is negative.
Their corresponding models are thus unphysical.
These past configurations and 
the best-fit straight lines of scaling and spacing rules 
are shown in Fig. 3 and Fig. 4, respectively.

According to Table 3, the values of both $\chi^2_{r1}$ and  $\chi^2_{r2}$
are smaller or around two for  
System 2, 3, 5, 6, 7, 8, 9. Indeed, as shown 
in Fig. 3-4, these systems 
follow both scaling and spacing rules and can be regarded
as the multi-planet systems which have AMD-model phases. 
 
We also plot the comparison of the semi-major axes  and the masses 
between the observational data and the past configurations 
in Fig. 5 and Fig. 6, respectively. 
The deviations caused by the orbital migrations and mass accretions 
are clearly presented.\\

{\centerline {\bf Table 3}}
\begin{center}
\begin{tabular}{|c|c|c|c|c|c|c|}\hline
ID & $p$  & $\chi^2_{r1}$ & $D_0$ &  $a_0^{b}$ & $a_0$ & $\chi^2_{r2}$ \\ 
   &   &         &(AU$^b$)&(AU$^b$)& (AU) &          \\ \hline 
1& $0.51_{-0.16}^{+0.16}$ & 3.17 & $0.32_{-0.01}^{+0.01}$ & 
$-0.17_{-0.016}^{+0.016}$ & -- & 21.53 \\ \hline
2& $0.53_{-0.31}^{+0.31}$ & $9.16\times 10^{-3}$ & 
$0.15_{-0.03}^{+0.03}$ & $0.041_{-0.056}^{+0.056}$ &  0.009 &0.623  \\ \hline
3 & $-0.72_{-0.95}^{+0.95}$ & $2.18\times 10^{-3}$ & $0.27_{-0.31}^{+0.31}$ &
$0.39_{-0.37}^{+0.36}$ & 0.027 & $9.76\times 10^{-3}$ \\ \hline
4 &$-4.90_{-0.12}^{+0.12}$ & 447.84& $4.04_{-0.30}^{+0.30}$ &
$21.32_{-1.34}^{+1.33}$ & 0.067 &5.80 \\ \hline
5 &$-0.79_{-0.24}^{+0.23}$ & 0.057  &  $0.66_{-0.18}^{+0.18}$ &
$0.72_{-0.087}^{+0.087}$ & 0.241 & 2.17 \\ \hline
6 & $-0.56_{-3.42}^{+3.38}$ & $1.49 \times 10^{-3}$ &
$0.14_{-1.78}^{+1.78}$ & $0.35_{-1.57}^{+1.58}$ & 0.037 & 
$2.68\times 10^{-4}$ \\ \hline
7 &$-0.07_{-0.39}^{+0.39}$ & $8.68\times 10^{-3}$ &$0.39_{-0.05}^{+0.05}$ & 
$0.44_{-0.08}^{+0.08}$ & 0.178 & 1.54  \\ \hline
8 &$-0.008_{-0.48}^{+0.46}$ & $0.055$ & $0.263_{-0.04}^{+0.04}$ &
$0.005_{-0.09}^{+0.09}$ &  $3.18\times 10^{-5}$ & 1.05\\ \hline
9 &$-2.00_{-0.99}^{+0.99}$ &  $1.47 \times 10^{-3}$ & 
$0.51_{-1.67}^{+1.66}$ & $1.50_{-0.96}^{+0.97}$ & 0.090 & 
$3.83\times 10^{-3}$ \\  \hline
10 & $3.26_{-0.84}^{+1.40}$ & 4.46 & $0.0057_{-0.0062}^{+0.0061}$ &
$-0.0066_{-0.012}^{+0.012}$ & -- & 1.87 \\ \hline
\end{tabular}
\end{center}

\section{Conclusions}

The origin of our Solar System is one of the most important
topics in astronomy. 
The existence of extra-solar multi-planet systems
has given a new opportunity
to understand the formation of planetary systems in general.
Since the AMD model is the only theory which gives
analytic expressions for both the mass and spacing distributions of 
planets, it is essential to know whether the detected extra-solar 
systems follow the AMD model. 

In this paper, the masses and semi-major axes of planets 
in ten multi-planet systems are taken to be examined.
It is found that four systems, i.e. GJ 667C, HD 215152, HD 40307,
and Kepler-79, follow the scaling and spacing rules of 
the AMD model.

Moreover, for the gas-poor scenario in which 
the mass accretion onto planets from the gaseous disk and
the orbital migrations caused by the gas were taken into account,
we find that GJ 3293, HD 141399, and HD 34445 systems 
were also close to the AMD-model configurations
and got scattered away later.

The embryos and planetesimals of the above seven systems were 
probably mainly grow is a gas-free or gas-poor environment.
The limited gaseous effects explain 
why these seven systems are consistent with the AMD model.

However, if the embryos and planetesimals grow in  
gas-rich environments, they would not be close to AMD model.
This could be why the configurations of other three systems,
i.e. 55 Cnc, GJ 876, and WASP-47 systems do not follow the AMD model.
Note that some of  these three systems, 
for example, 55Cnc, could still follow the TBR 
(Cuntz 2012), which could be driven
by other unrelated mechanisms. 
To conclude, our results provide important information 
on the formation histories of these multi-planet systems.

\section*{Acknowledgments}
We are grateful to the referee, Manfred Cuntz, 
for very helpful suggestions.
This work is supported in part 
by the Ministry of Science and Technology, Taiwan, under
Li-Chin Yeh's Grant MOST106-2115-M-007-014
and Ing-Guey Jiang's Grant MOST106-2112-M-007-006-MY3.

\clearpage
\begin{figure}
\epsscale{1.1}
\plotone{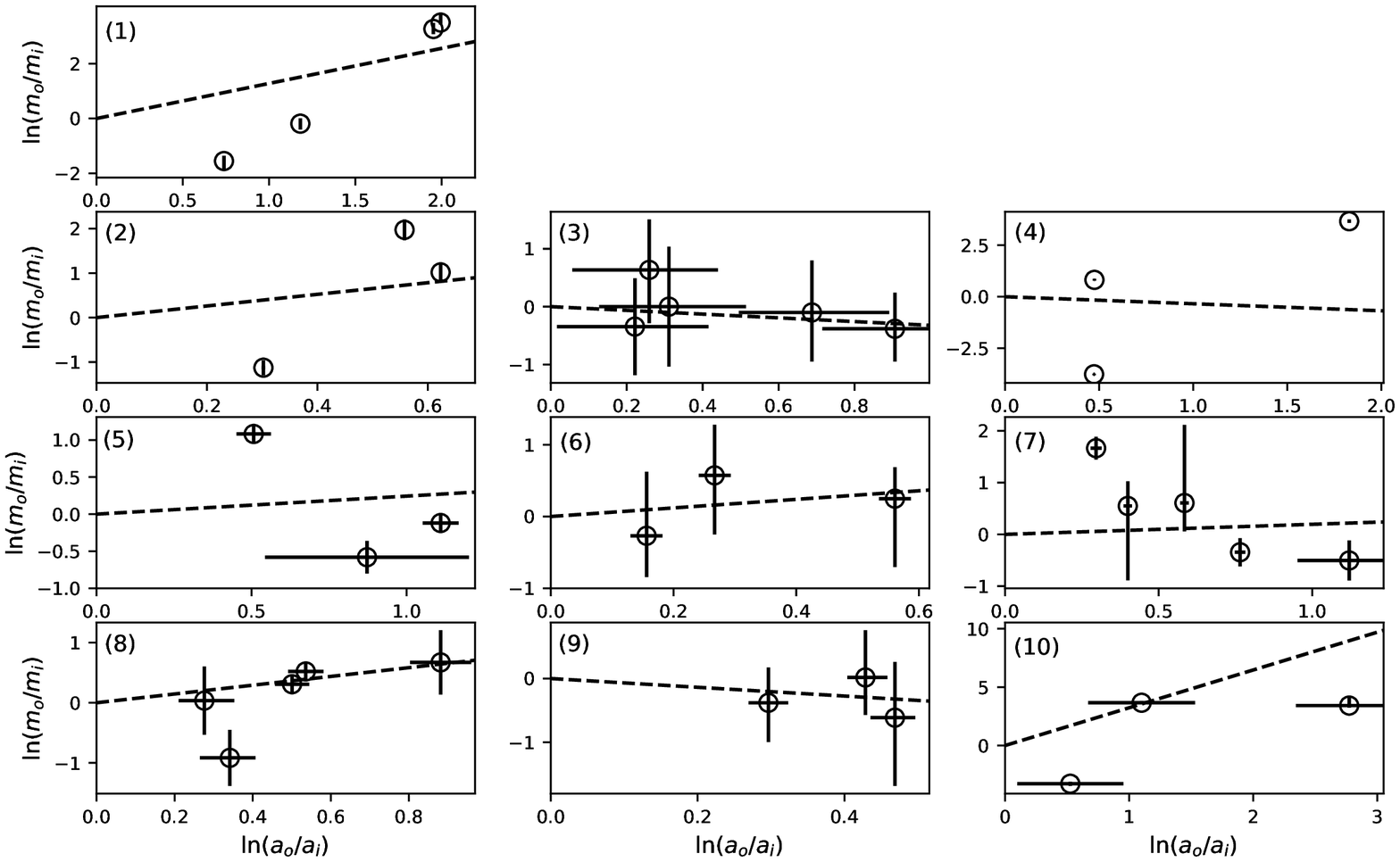}
\caption{The scaling law with the current data. 
The circles are the observational data in the 
$\ln (a_o/a_i)$-$\ln (m_o/m_i)$ plane.
The error bars are determined by equations in Appendix B. 
The straight dashed lines are the best-fit models.   
Panel (i) is for System i, where i=1,2,...,10.
\label{fig1}}
\end{figure}

\begin{figure}
\epsscale{1.1}
\plotone{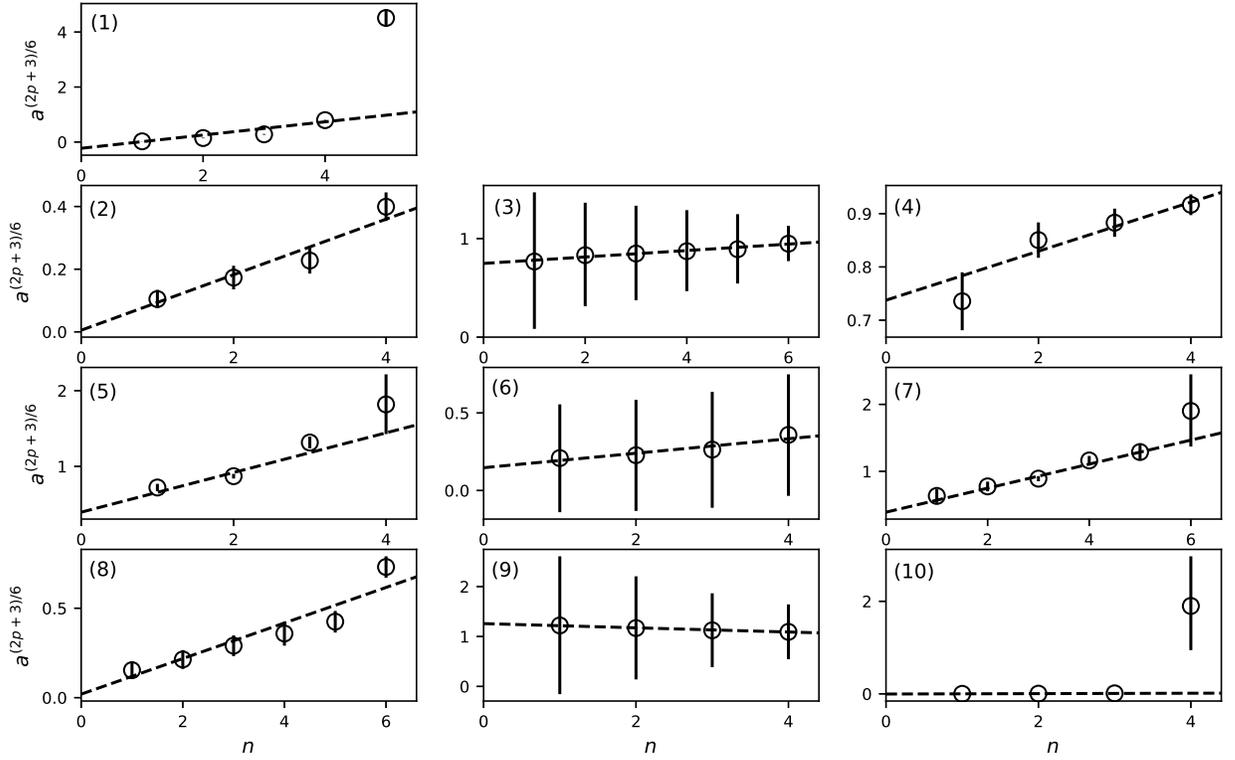}
\caption{The spacing law with the current data. 
The circles are the observational data 
in the $n$-$a^{(2p+3)/6}$ plane, where $n$ is the planet order.  
The error bars are determined by equations in Appendix B.
The straight dashed lines are the best-fit models.  
Panel (i) is for System i, where i=1,2,...,10.
\label{fig2}}
\end{figure}  

\begin{figure}
\epsscale{1.1}
\plotone{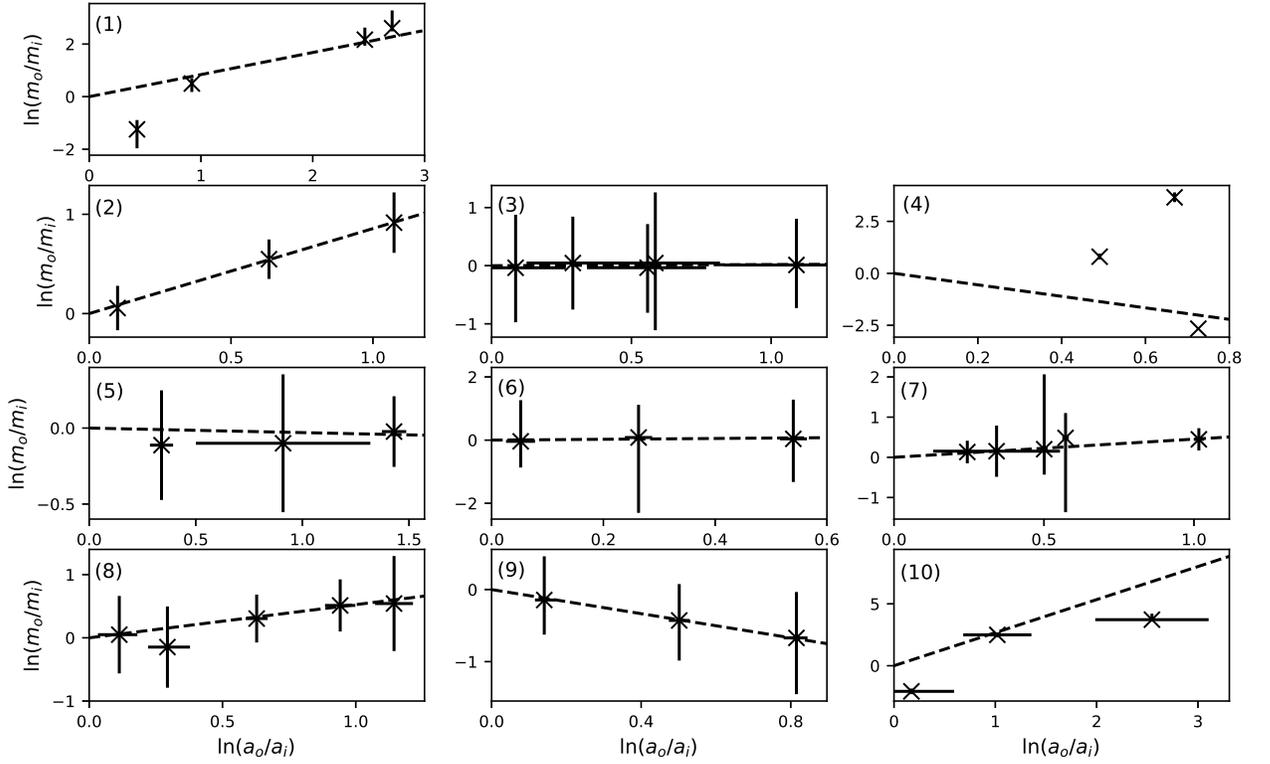}
\caption{The scaling law with hypothetical data. 
The crosses are the data in the $\ln (a_o/a_i)$-$\ln (m_o/m_i)$ plane. 
The error bars are determined by equations in Appendix B.
The straight dashed lines are the best-fit models. 
Panel (i) is for System i, where i=1,2,...,10.
\label{fig3}}
\end{figure}
 
\begin{figure}
\epsscale{1.1}
\plotone{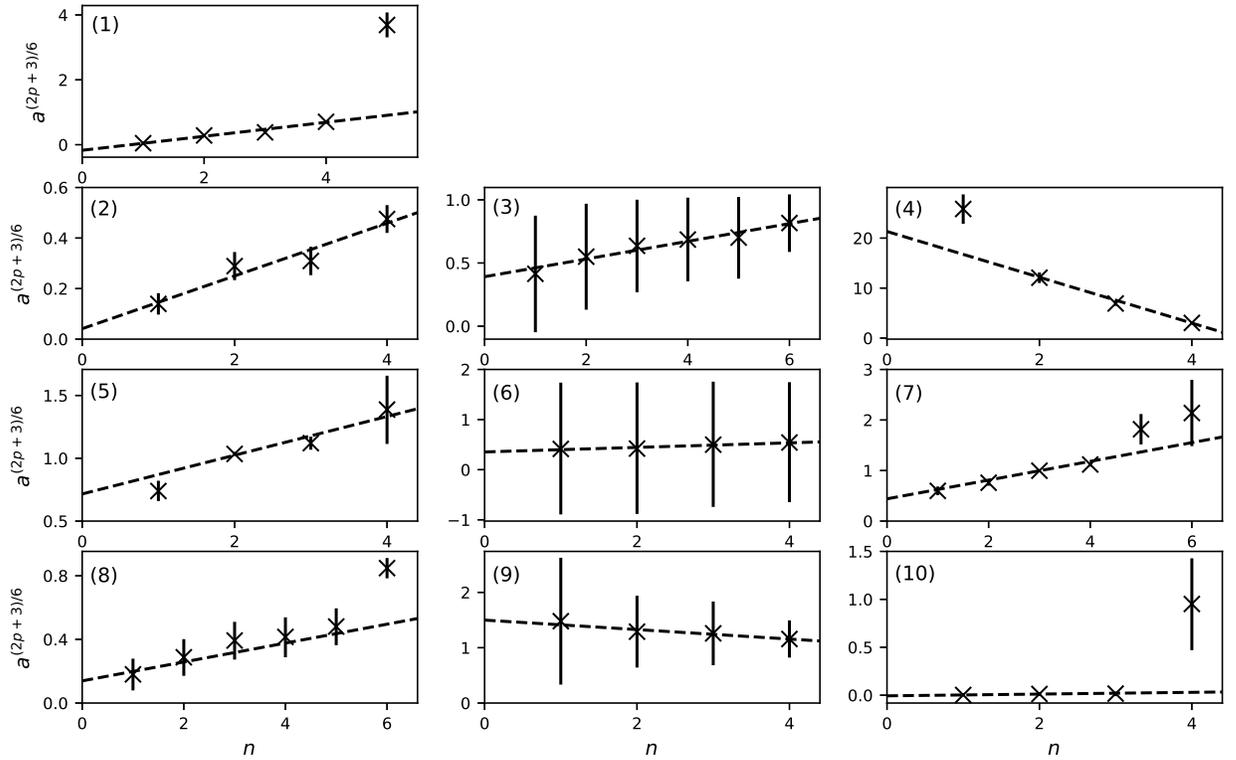}
\caption{The spacing law with hypothetical data. 
The crosses are the data in the $n$-$a^{(2p+3)/6}$ plane, 
where $n$ is the planet order. 
The error bars are determined by equations in Appendix B.
The straight dashed lines are the 
best-fit models. Panel (i) is for System i, where i=1,2,...,10.
\label{fig4}}
\end{figure} 

\begin{figure}
\epsscale{1.1}
\plotone{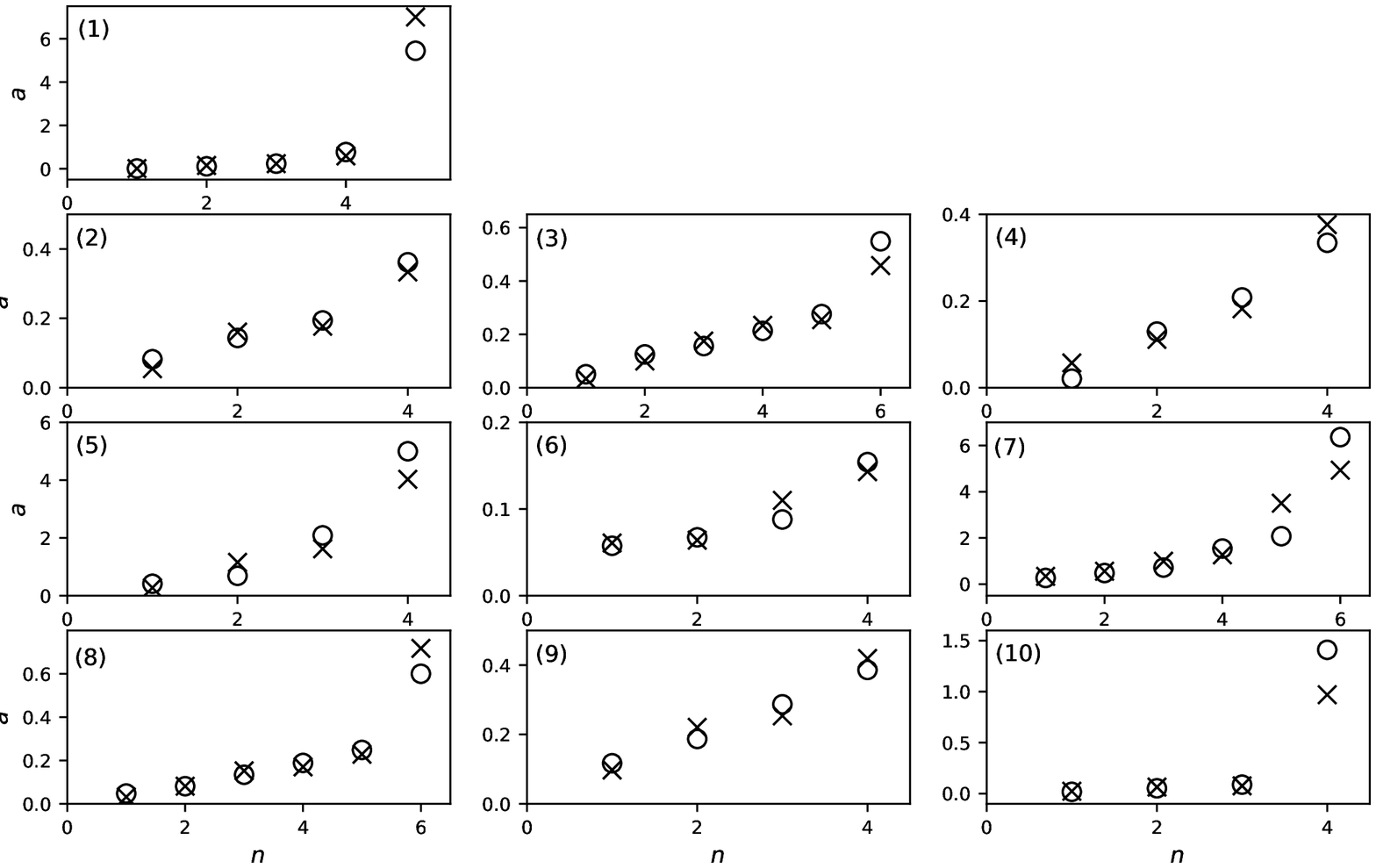}
\caption{The observed and hypothetical 
values of semi-major axes of all planets.
The circles are the observational and crosses are
the hypothetical.  
Panel (i) is for System i, where i=1,2,...,10.
\label{fig5}}
\end{figure} 

\begin{figure}
\epsscale{1.1}
\plotone{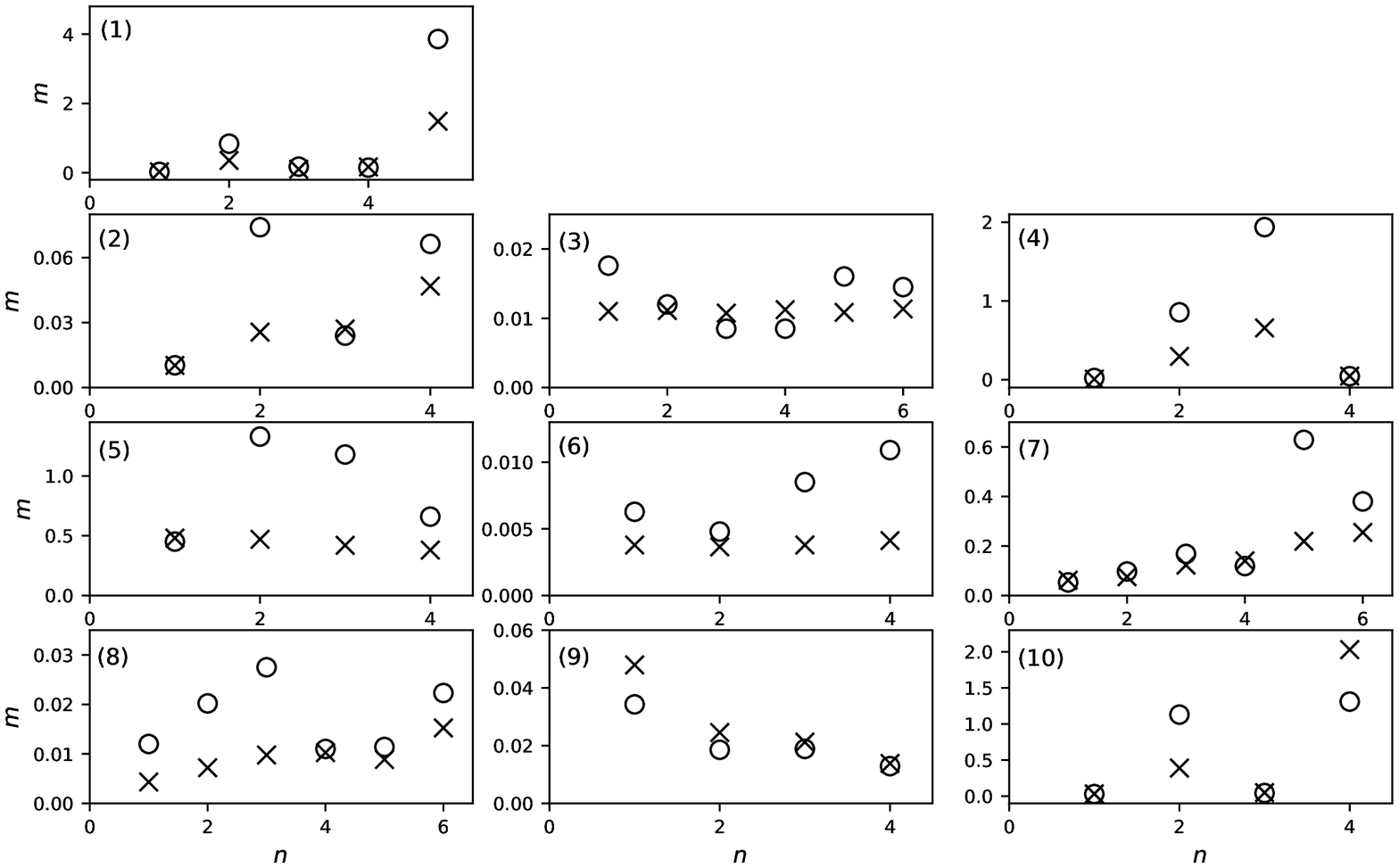}
\caption{The observed and hypothetical 
values of masses of all planets.
The circles are the observational and crosses are
the hypothetical.  
Panel (i) is for System i, where i=1,2,...,10.
\label{fig6}}
\end{figure} 

\clearpage
{\center{\bf Appendix A} \\
\vspace {0.2 in}
}
The data of ten systems are listed here, where
$m$ is the planetary minimum mass with lower error bound $m_{-}$
and  upper error bound $m_{+}$;
$a$ is the planetary semi-major axis 
with lower error bound $a_{-}$
and  upper error bound $a_{+}$.

\begin{tabular}{|c|c|c|c|c|c|c|c|}\hline
ID & Name & $m$ & $m_{-}$ & $m_{+}$ & $a$ &$ a_{-}$ & $a_{+}$ \\ 
   &      & ($M_J$)&($M_J$)&($M_J$)& (AU) & (AU) & (AU) \\\hline
\multicolumn{1}{|c|}{1}
& 55 Cnc b & 0.84 &   0.031&   0.23 &   0.11339& 0.00011 & 0.00011 \\
& 55 Cnc c  &0.1784 & 0.0078 &  0.0275 &  0.23735 & 0.00024 & 0.00024 \\
& 55 Cnc d  & 3.86  &  0.15  &  0.6  &   5.446 &   0.02 &   0.02  \\
& 55 Cnc e  &0.02547 & 0.00081& 0.00089& 0.015439 & $1.50E^{-5}$ & $1.50E^{-5}$\\
& 55 Cnc f  & 0.1479 & 0.0093 & 0.0219 & 0.7733 & 0.001 &  0.001\\ \hline
\multicolumn{1}{|c|}{2}
&GJ 3293 b & 0.07406  &0.0028  &0.0028  &0.14339 &0.0003  &0.0003 \\
&GJ 3293 c & 0.06636  &0.00396 &0.00396 &0.36175 &0.00048 &0.00048 \\
&GJ 3293 d & 0.024  & 0.0031  &0.0031  &0.194   &0.00018 &0.00018 \\
&GJ 3293 e & 0.0103 & 0.002   &0.002   &0.0821  &$4.00E^{-5}$ & $4.00E^{-5}$ \\
 \hline
\multicolumn{1}{|c|}{3}
&GJ 667C b  &   0.0176 & 0.0041 & 0.0044 & 0.0505 & 0.0053 & 0.0044\\
&GJ 667C c  &   0.012  &   0.0038 &  0.0047 & 0.125 &  0.013 &  0.012\\
&GJ 667C d  &   0.01604 & 0.00535& 0.00566 & 0.276 &  0.03  &  0.024\\
&GJ 667C e  &   0.0085 & 0.0044 & 0.005  & 0.213  & 0.02   & 0.02\\
&GJ 667C f  &   0.0085 & 0.0038 & 0.0044 & 0.156  & 0.017  & 0.014\\
&GJ 667C g  &   0.0145 & 0.0072 & 0.0082 & 0.549  & 0.058  & 0.052\\ \hline
\multicolumn{1}{|c|}{4}
&GJ 876 b &1.938&   0.014&   0.036&   0.208317 & $2.00E^{-5}$ &  $2.00E^{-5}$\\
&GJ 876 c &0.856&   0.029&   0.032&   0.12959 &$2.40E^{-5}$    & $2.40E^{-5}$ \\
&GJ 876 d &0.022&   0.001&   0.001&   0.02080665& $1.50E^{-7}$ & $1.50E^{-7}$ \\
&GJ 876 e &0.045&   0.001&   0.001&   0.3343  &0.0013 &  0.0013 \\ \hline
\multicolumn{1}{|c|}{5}
&HD 141399 b  &  0.451 & 0.03 &  0.03 &   0.415 &  0.011 &  0.011 \\
&HD 141399 c  &  1.33  & 0.08 &  0.08 &   0.689 &  0.02  &  0.02 \\
&HD 141399 d  &  1.18  & 0.08 &  0.08 &   2.09  &  0.06  &  0.06 \\
&HD 141399 e  &  0.66  & 0.1  &  0.1  &   5     &  1.5   &  1.5  \\ \hline
\multicolumn{1}{|c|}{6}
& HD 215152 b  &  0.00628 & 0.00235 & 0.0013  & 0.057635 &0.000755 &   
 0.000742 \\
& HD 215152 c  &   0.004801 &0.00177 & 0.00249 & 0.067399 & 0.000898 & 
 0.000854 \\
&HD 215152 d & 0.008511 & 0.00259 & 0.00287 & 0.08799 &0.00115 &0.00113 \\
&HD 215152 e & 0.01091 &0.006711  & 0.00145 & 0.15417 &0.00204 &0.00199 \\ 
\hline
\end{tabular}
\clearpage
\begin{tabular}{|c|c|c|c|c|c|c|c|}\hline
ID & Name & $m$ &$m_{-}$ & $m_{+}$ & $a$ &$a_{-}$ & $a_{+}$ \\ 
   &      & ($M_J$)&($M_J$)&($M_J$)& (AU) & (AU) & (AU) \\ \hline
\multicolumn{1}{|c|}{7}
&HD 34445 b &  0.629 &  0.028 &  0.028 &  2.075 &   0.016 &  0.016 \\
&HD 34445 c &  0.168 &  0.016 &  0.016 &  0.7181 &  0.0049 & 0.0049 \\
&HD 34445 d &  0.097 &  0.037 &  0.13  &  0.4817 & 0.0033 & 0.0033 \\
&HD 34445 e &  0.0529 &  0.0089 & 0.0089 & 0.2687 & 0.0019 & 0.0019 \\
&HD 34445 f &  0.119  & 0.021 &  0.021 &  1.543  & 0.016  & 0.016 \\
&HD 34445 g &  0.38   & 0.13  &  0.13  &  6.36   & 1.02  &  1.02 \\ 
\hline
\multicolumn{1}{|c|}{8}
&HD 40307 b &  0.012 &  0.00094 &0.00094 &0.0475 & 0.0011 & 0.0011 \\
&HD 40307 c &  0.0202 & 0.0014  &0.0014  &0.0812 & 0.0018 & 0.0018 \\
&HD 40307 d &  0.0275 & 0.0018  &0.0018  &0.134  & 0.0029 & 0.0029 \\
&HD 40307 e &  0.011  & 0.0044  &0.0044  &0.1886 & 0.0104 & 0.0083  \\
&HD 40307 f &  0.0114 & 0.0019  &0.0019  &0.2485 & 0.0054 & 0.0054  \\
&HD 40307 g &  0.0223 & 0.0082  &0.0082  &0.6    & 0.034  & 0.034 \\ \hline
\multicolumn{1}{|c|}{9}
&Kepler-79 b& 0.034285 & 0.018872   & 0.023276 &0.117  & 0.002  & 0.002 \\
&Kepler-79 c& 0.018558 & 0.007234  &  0.005976 & 0.187 &  0.003 &  0.002 \\
&Kepler-79 d& 0.018872 & 0.005033  &  0.006605 & 0.287 &  0.004 &  0.004 \\
&Kepler-79 e& 0.0129 & 0.0034 &  0.0037 & 0.386 &  0.005 &  0.005 \\ \hline
\multicolumn{1}{|c|}{10}
&WASP-47 b &  1.13 &   0.038 &   0.038 &  0.052 &   0.011&   0.011 \\
&WASP-47 c &  1.31 &   0.05  &  0.72  &  1.41  &  0.3  &   0.3 \\
&WASP-47 d &  0.0428 &  0.0063 &  0.0063 & 0.088 &   0.019 &  0.019 \\
&WASP-47 e &  0.029 &  0.0031 & 0.0031 & 0.0173 & 0.0038 & 0.0038 \\ \hline
\end{tabular}

\clearpage
{\center{\bf Appendix B} \\
}
\vspace {0.2 in}
\no {\bf (1)}

Let the parameter $p=X^{+x_{+}}_{-x_{-}}$, i.e. a value $X$ 
with upper error bound $x_{+}$  and  lower error bound $x_{-}$; 
the parameter $q=Y^{+y_{+}}_{-y_{-}}$, i.e. 
a value $Y$ with upper error bound $y_{+}$ 
and lower error bound $y_{-}$.
In order to determine the upper and lower error bounds
of the parameter ratio $p/q$, we need the expression 
that when $|s| << 1$, ${1}/{(1+s)} \approx 1-s$.
From below
$$\frac{p}{q}=\frac{X^{+x_{+}}_{-x_{-}}}{Y^{+y_{+}}_{-y_{-}}},$$
we know the maximum is
\beqn
\frac{p}{q}&=&\frac{X+x_{+}}{Y-y_{-}}
=\frac{X}{Y}\frac{\left(1+\frac{x_{+}}{X}\right)}
      {\left(1-\frac{y_{-}}{Y}\right)}
\approx \frac{X}{Y}\left(1+\frac{x_{+}}{X}
\right) \left(1+\frac{y_{-}}{Y}\right) \non \\
&\approx&\frac{X}{Y}\left(1+\frac{x_{+}}{X}+\frac{y_{-}}{Y}\right),
\eeqn
and the upper error bound is 
\beq
\frac{X}{Y}\left(\frac{x_{+}}{X}+\frac{y_{-}}{Y}\right).
\eeq
Similarly, the minimum is 
\beqn
\frac{p}{q}&=&\frac{X-x_{-}}{Y+y_{+}}
=\frac{X}{Y}\frac{\left(1-\frac{x_{-}}{X}\right)}
      {\left(1+\frac{y_{+}}{Y}\right)}
\approx \frac{X}{Y}\left(1-\frac{x_{-}}{X}
\right) \left(1-\frac{y_{+}}{Y}\right) \non \\
&\approx&\frac{X}{Y}\left(1-\frac{x_{-}}{X}-\frac{y_{+}}{Y}\right),
\eeqn
and the lower error bound is 
\beq
\frac{X}{Y}\left(\frac{x_{-}}{X}+\frac{y_{+}}{Y}\right).
\eeq

\vspace{0.1 in} 
\no {\bf (2)}

Given that the value of $x$ has an upper error bound $x_{+}$ and a 
lower error bound $x_{-}$, 
the upper and lower error bounds
of a general function $f(x)$ can be determined  
through Taylor series. 
When $x_{+}$ and $x_{-}$ are small, using 
$\Delta x$ to denote $x_{+}$ or $-x_{-}$,
we have
\beq
f(x+\Delta x)\approx f(x)+f'(x)\Delta x +{\rm O}((\Delta x)^2).
\label{eq:fx+}
\eeq
The term $f'(x)\Delta x$ is thus used as the error estimation.
It equals to $f'(x)x_{+}$ or $-f'(x)x_{-}$.
When $f'(x)$ is positive, 
the upper error bound of $f(x)$ is $f'(x)x_{+},$ 
and the lower error bound of $f(x)$ is $f'(x)x_{-}.$
When $f'(x)$ is negative, 
the upper error bound of $f(x)$ is $-f'(x)x_{-},$ 
and the lower error bound of $f(x)$ is $-f'(x)x_{+}.$
For example, when the function  $f(x)=\ln(x)$, 
$f'(x)>0$, 
the upper and lower error bounds  
are ${x_{+}}/{x}$ and ${x_{-}}/{x}$.

\vspace{0.1 in} 
\no {\bf (3)}

Given that the value of $x$ has an upper error bound $x_{+}$ and a 
lower error bound $x_{-}$, 
the value of $y$ has an upper error bound $y_{+}$ and a 
lower error bound $y_{-}$, 
the upper and lower error bounds
of a general function $f(x,y)$, where $x$ and $y$ are two independent
variables, can be determined through Taylor series. 
When $x_{+}$, $x_{-}$, $y_{+}$, and $y_{-}$ are small, using 
$\Delta x$ to denote $x_{+}$ or $-x_{-}$ 
and $\Delta y$ to denote $y_{+}$ or $-y_{-}$,
we have
\beq
f(x+\Delta x, y+\Delta y)\approx f(x,y)
+\frac{\partial f(x,y)}{\partial x}\Delta x 
+\frac{\partial f(x,y)}{\partial y}\Delta y 
+{\rm O}( (\Delta x)^2, (\Delta y)^2, (\Delta x)(\Delta y) ).
\label{eq:fxy+}
\eeq
The summation of below two terms, i.e. 
\beq
\frac{\partial f(x,y)}{\partial x}\Delta x 
+\frac{\partial f(x,y)}{\partial y}\Delta y, 
\eeq
is thus used as the error estimation.
Its four possible values can be determined after
$\Delta x$ is substituted by $x_{+}$ or $-x_{-}$, 
and $\Delta y$ is substituted by $y_{+}$ or $-y_{-}$. 
The maximum value would be positive, and the minimum value would
be negative. 
The upper error bound of $f(x,y)$ is then equal to this maximum,
and the lower error bound of $f(x,y)$ is the absolute value of the
minimum.

\end{document}